\documentclass[aps,pre,twocolumn]{revtex4}
\usepackage{graphicx}
\usepackage{mathptmx}
\usepackage{amsmath, amssymb}
\usepackage{bm} 
\usepackage{color}

\usepackage{amsfonts,amsmath,amssymb,amsthm}
\def\beq{\begin{equation}}
\def\eeq{\end{equation}}
\def\bea{\begin{eqnarray}}
\def\eea{\end{eqnarray}}
\usepackage[normalem]{ulem}
\usepackage{cancel}

\usepackage{hyperref}
\begin{document}
\makeatletter
\title{Active chain spirograph: Dynamic patterns formed in extensible chains due to follower 
activity}

\author{Sattwik Sadhu}
\email{sat.sadhrsb@gmail.com}
\author{Nitin Kriplani}
\author{Anirban Sain}
\author{Raghunath Chelakkot}
\email{raghu@phy.iitb.ac.in}

\affiliation{Department of Physics,
Indian Institute of Technology
Bombay, Powai, Mumbai--400 076, India.}

\begin{abstract}
\noindent
{Follower activity results in a large variety of conformational and dynamical states in active 
chains and filaments. These states are formed due to the coupling between chain geometry and the local activity. We 
study the origin and emergence of such patterns in noiseless, flexible active chains. In the overdamped limit, we
observed a range of dynamical steady states for different chain lengths ($N$). The steady-state planar trajectories of the centre-of-mass of the chain include circles, periodic waves, and quasiperiodic, bound trajectories resembling spirographic patterns. In addition, out-of-plane initial configuration also leads to the formation of 3D structures, including globular and supercoiled helical structures. For the shortest chain with three segments $(N=3)$, the chain always moves in a circular trajectory. Such circular trajectories are also observed in the limit of large chain lengths $(N \gg 1)$. We analytically study the dynamical patterns in these two limiting cases, which show quantitative and qualitative matches with numerical simulations. 
Our analytical study also provides an estimate of the limiting $N$ where the large chain length behaviour is expected. These analyses reveal the existence of such intricately periodic patterns in active chains, arising due to the follower activity.  
}
\end{abstract}

\maketitle

\section{Introduction}
 Understanding the collective dynamics of self-propelled particles continues to be an interesting topic across multiple disciplines. One of the characteristic properties of such active systems is the rich dynamical patterns generated by collectively moving agents, which cannot be predicted solely from their individual dynamics~\cite{vicsek2012collective}. These patterns include the formation of swarms~\cite{attanasi2014finite, patel2022formation, kelley2013emergent}, flocks~\cite{bialek2012statistical,cavagna2014bird,attanasi2015emergence}, lanes~\cite{feliciani2016empirical, bacik2023lane, couzin2003self, murakami2019levy}, vortices~\cite{franks2016social, bazazi2012vortex, delcourt2016collective} , and travelling waves~\cite{welch2001cell, allard2013traveling}, and they are found in a variety of biological and inanimate active systems. Such emergent dynamical patterns exhibited by active agents under varying conditions are influenced by factors such as activity, inter-particle interactions, and various environmental factors~\cite{marchetti2013hydrodynamics, bechinger2016active}.

The interaction between active agents is one of the crucial factors in determining the type of patterns they form collectively. In certain artificial active systems such as vibrating granular rods, the inter-particle interactions manifest as reciprocal forces, which can be derived from an interaction potential~\cite{narayan2007long, arora2024shape, aranson2007swirling}. However, in another class of active systems, inter-particle interactions are more complex and inherently non-reciprocal. These include synthetic colloidal particles driven via phoretic and thermo-osmotic forces, as well as a broad range of living and biological systems across multiple length scales with various complex interactions~\cite{lavergne2019group, gomez2022intermittent, duan2023dynamical, you2020nonreciprocity, dinelli2023non, fruchart2021non}. In many of these systems, the interactions between active entities arise as each entity responds to the environmental cues generated by the others.

In theoretical studies of active manybody systems, the interactions are implemented as a combination of reciprocal and non-reciprocal interactions to study the collective behaviour of the active systems. For example, a large class of particle-based models implement pairwise interactions to impose a minimum distance between the particles. In addition to these passive interactions, minimal non-reciprocal interactions are also implemented~\cite{baconnier2025self, hiraiwa2020dynamic, das2024flocking}. Although these interactions are imposed as relatively simple rules for each particle, they mimic the qualitative nature of complex interactions observed in synthetic and biological systems and
 lead to novel collective patterns, which are not observed in active systems interacting only via reciprocal forces~\cite{baconnier2025self, hiraiwa2020dynamic} .

In addition to systems of self-motile particles, which propel without any geometric constraints, a large number of studies have also been conducted on the behaviour of active chains and filaments, made of active segments that are physically or functionally linked~\cite{winkler2020physics}. A significant portion of these systems includes non-resiprocal interactions with their neighbours, via the mechanism called follower forces~\cite{chelakkot2014flagellar, de2017spontaneous, krishnamurthy2023emergent, elgeti2015physics, fily2020buckling, sangani2020elastohydrodynamical, Fatehiboroujeni2018, fatehiboroujeni2021three,thakur2022self, jain2022cargo, Laskar2013}. In this follower force mechanism, the self-propulsion direction of each segment is preferentially aligned along the local tangent of the chain, which is determined by the location of the neighbouring segments. 
Since they are geometrically constrained, the follower interaction between the segments gives rise to a highly coordinated, collective behaviour, depending on the strength of propulsion, elastic stiffness, and the 
boundary conditions~\cite{chelakkot2014flagellar, elgeti2015physics, Fatehiboroujeni2018, fazelzadeh2023effects, anand2018structure,ghosh2014dynamics, isele2015self, peterson2020statistical,prathyusha2018dynamically, karan2024inertia, anand2019beating, kharayat2025kinetically}. 
\
\
A key feature observed in connected active systems with non-zero bending rigidity and under follower activity is the emergence of periodic oscillations and rotations, when one end of the chain is clamped or pivoted to a rigid base~\cite{chelakkot2014flagellar, de2017spontaneous, ling2018instability, Fatehiboroujeni2018, fily2020buckling, sangani2020elastohydrodynamical}. Such periodic motions share striking similarity to the dynamics of Eukaryotic cilia and flagella. In addition, in vitro systems containing intracellular filaments such as microtubules and actin, driven by molecular motors, have also shown periodic oscillations of filaments~\cite{bar2020self, Sanchez2011, yadav2024wave, sen2024coordination,collesano2022active, shee2021semiflexible}. In such microscopic elastic systems, the bending rigidity of the filaments suppresses the lateral fluctuations in the filament and the local tangent direction, thereby facilitating a persistent activity that enables coordinated motion. In the absence of bending rigidity, flexible active polymer models with follower force have shown non-equilibrium conformational changes without any periodic motion, as expected due to a significant amount of thermal noise~\cite{bianco2018globulelike, li2024activity, jaiswal2024diffusiophoretic,li2023nonequilibrium, wang2022conformation, malgaretti2025coil, fazelzadeh2023effects}. This raises an intriguing question about the possibility of finding periodic motion even in a flexible chain with no bending rigidity, if the noise is set to zero. While noise is inherent to microscopic systems, these questions are relevant given the growing interest in autonomous locomotive systems and their technological applications~\cite{riedl2023synchronization, paramanick2024uncovering, son2025emergent}.

In this work, we systematically study the dynamical states of a flexible chain of active agents in the noiseless limit. We analytically show that the tangential follower activity leads to a periodic motion even for the smallest chain of three monomers. For chains of intermediate length, the numerical studies reveal surprisingly rich dynamical trajectories of the centre-of-mass of the chain, depending on the chain length and initial conditions. The bound centre-of-mass trajectories resemble spirograph patterns, and the unbound trajectories show complex wave-like patterns. When the chain length is sufficiently large, the chain assumes a circular configuration and the centre-of-mass follows a circular trajectory. 

\
\
\

\section{Model and methods}

\begin{figure}[h]
    \centering
\includegraphics[width=.5\textwidth]{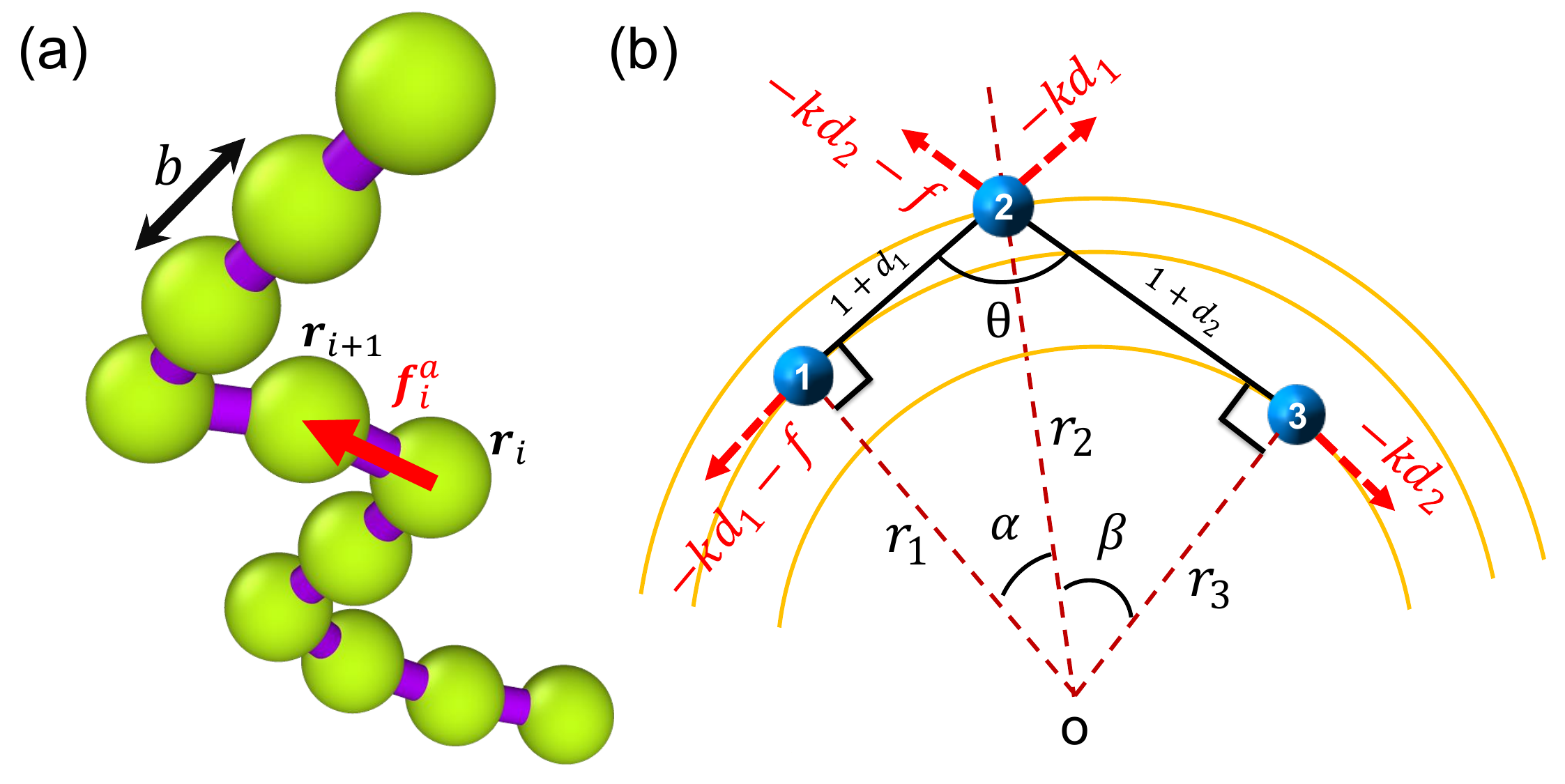}
    \caption{(a) Schematic of the active chain showing the activity scheme used. (b) Schematic showing the forces on different monomers for a three bead chain.}
    \label{fig1}
\end{figure}

The minimal active chain model consists of $N$ particles (monomers), successively connected by linear springs, forming a flexible chain. Each monomer $i$ is connected to two of its neighbours via a  elastic potential, given by  $V(\bm{r}_1,\dots,\bm{r}_N)=\frac{1}{2}k\sum_{i=1}^{N-1}(\lvert\bm{r}_{i+1}-\bm{r}_i\rvert-b)^2$, where $\bm{r}_i$ denotes the position of the $i^{th}$ particle, $b$ the equilibrium length and $k$ the force constant of the connecting springs. In addition, an active force $\bm {f}_i^a = f\hat{\bm{p}}_i$ is applied on each monomer, where $\hat{\bm{p}_i}$ is the unit tangent vector defined for the $i^{th}$ monomer. In this study, we mainly focus on systems where the unit tangent vector is defined as $\hat{\bm{p}_i}=\hat{\bm{\Delta}}_i= (\bm{r}_{i+1} - \bm{r}_i)/|\bm{r}_{i+1} - \bm{r}_i|$. The activity scheme is shown in fig. \ref{fig1}(a).

Thus, the flexible chain we study consists of monomers with polar activity and each of the monomers driven in the direction of the instantaneous tangent, thus coupling the local activity to the geometry of the chain. According to this definition of activity, the leading monomer ($i=N$) remains passive.

We study the system in the overdamped limit by numerically integrating the equation $\dot{\bm{r}}_i(t)=\gamma^{-1}(\bm{F}_i(t)+\bm{f}^a_i(t))$, where $\bm{F}_i=-\bm{\nabla}_iV$ is the passive conservative component of the force due to harmonic springs and $\bm{f}^a_i$ is the active force as mentioned above. In addition, we verify all the results by simulating the inertial system represented by the equation $m\ddot{\bm{r}}_i(t)+\gamma\dot{\bm{r}}_i(t)=\bm{F}_i(t)+\bm{f}^a_i(t)$ in high damping limit $\gamma \gg 1$. For the parameter values, we take $b=m=1$, $k=100\gamma$ and $f/k=1$ unless specified otherwise.
The overdamped equation is numerically solved using the fourth-order Runge-Kutta (RK4) Scheme, and for the underdamped simulations, we used both a second-order Velocity Verlet scheme and RK4 scheme to check the consistency of the obtained results.  The timestep $\Delta t$ is chosen to be in the range $10^{-3}$ to $10^{-5}$ which was adjusted according to the parameter values to ensure consistency and convergence. The numerical studies were conducted for values of $N$ ranging from $3$ to $200$.
\

\section{results}
We have numerically studied the system for a range of parameters. We observed that there are different kinds of steady state dynamics in our system as we vary $N$ in the overdamped limit. 
To categorise these steady states, we examine the temporal evolution of the configurational parameters such as bond-lengths $\Delta_i = {|{\bf r}_{i+1} - {\bf r}_i|}$, and bond angles $\theta_i=\cos ^{-1}(\bm{\hat\Delta}_i \cdot \bm{\hat\Delta}_{i+1})$, $\bm{\hat \Delta_i}$ being the unit bond vector, and also dynamical quantities like the centre-of-mass (COM) of the entire chain. If $\Delta_i$ and $\theta_i$ are constant in time ($\dot{\theta}_i = \dot{\Delta}_i =0$), we identify the system to be in a `rigid' steady state. Conversely, if $\dot{\Delta}_i$ and $\dot{\theta}_i$ are non-zero, the system is in a `flexible' steady-state. In addition, depending on the parameters, the COM may get confined in a finite region in space, leading to a `bound' steady state, or shift continuously corresponding to an `unbound' steady state. Thus, the two defining properties of the steady state, namely the time-evolution of the configurational parameters and the spatial localisation of the COM are utilized to classify the steady states. For example, the trivial case of a straight configuration uniformly translating in space is a rigid, unbound state. Similarly, we obtain other flexible and rigid states, both bound and unbound, as detailed in the subsequent sections.
\subsection{Analysis of rigid states}
We first analyze the rigid states analytically, for which the bond lengths $\Delta_i$ and angles $\theta_i$ remain fixed over time, although the orientation of a rigid polymer can change. In addition, rigidity also guarantees that the angle between the chain orientation and the net force is constant and the direction of force reorients with the same rate as the chain. If this angle is non-zero, this  ensures a circular trajectory for each monomers. To analyze the rigid state, we first express the internal spring force on each monomer $i$ in the form $\bm{F}_i=k\{(\Delta_i-b)\hat{\bm{\Delta}}_i(1-\delta_{i,N})-(\Delta_{i-1}-b)\hat{\bm{\Delta}}_{i-1}(1-\delta_{i,1})\}$, where $\delta_{i,j}$ is the Kronecker delta function. 
Using this, we write the underdamped equation in the dimensionless form, first by replacing the position and time by their dimensionless counterparts as $\bm{r}_i\rightarrow b^{-1}\bm{r}_i$ and $t \rightarrow k\gamma^{-1}t$ respectively and subsequently by inroducing dimensionless force $\tilde{f}=\frac{f}{kb}$ and mass $\tilde{m}=\frac{mk}{\gamma^2}$. The non-dimensional equation reads,
\begin{equation}\label{eq1}
    \tilde{m}\ddot{\bm{r}}_i(t)+\dot{\bm{r}}_i(t)=(d_i+\tilde{f})\hat{\bm{\Delta}}_i(1-\delta_{iN})-d_{i-1}\hat{\bm{\Delta}}_{i-1}(1-\delta_{i1}).
\end{equation}
Here $d_i = \Delta_i -1$ is the extension of the $i^{\text{th}}$ bond.
Now that all the variables and parameters are in their dimensionless forms, the $\tilde{(\cdot)}$ will be dropped in the subsequent analyses, but they will continue to represent their dimensionless forms unless specified otherwise. The overdamped equation is obtained simply by setting $m=0$.
To analyze the rigid states, it is convenient to rewrite eq \eqref{eq1} in terms of the bond vector ${\bm \Delta}_i$ providing,
\begin{equation}\label{eq5}
m \ddot{\bm{\Delta}}_i + \dot{\bm{\Delta}}_i = (d_{i+1} + f ) \hat{\bm{\Delta}}_{i+1} - (2d_i + f ) \hat{\bm{\Delta}}_i + d_{i-1} \hat{\bm{\Delta}}_{i-1}
\end{equation}
for $i=1,\dots,N-1$ with $d_0=d_N+f=0$ (boundary conditions). 
Let us denote $\bm{\Delta}_i=\Delta_i(\cos \phi_i, \sin \phi_i)^T$, where $\phi$ is the angle made by $\bm{\Delta}$ with the $x$ axis. For rigid states,  $\dot{\Delta}_i=0$ and $\dot{\phi}_i=\omega$ for $i=1$ to $N-1$ by definition, where $\omega$ is the angular velocity of the polymer. Now, taking the inner product with respect to $\hat{\bm{\Delta}}_i$ and $\hat{\dot{\bm{\Delta}}}_i$ respectively on both sides of eq \eqref{eq5} and using the definition of rigidity and bond angle, we get
%
%
%
\begin{subequations}\label{eq7}
\begin{align}
(d_{i+1}+f) \cos \theta_i-(2d_i+f-m\omega^2(d_i+1))+d_{i-1} \cos \theta_{i-1}  & = 0,\label{eq7a}\\
\text{and}\qquad\qquad\;(d_{i+1}+f) \sin \theta_i-(d_i+1)\omega-d_{i-1} \sin \theta_{i-1} & =0\label{eq7b}
\end{align}
\end{subequations}
The relations given in \eqref{eq7} are valid for all rigid states and serve as the starting point for most of the subsequent analysis. Note that these equations have a lesser number of degrees of freedom compared to the original eq \eqref{eq1}. This is because we set $\dot{\Delta} =0$ and captured the `dynamics' part from eq \eqref{eq1}, by a single variable $\omega$, effectively emphasizing only the configurational information, from which the dynamics can easily be recovered for a rigid state.  Setting mass $m=0$ in eq \eqref{eq7}, we get the corresponding overdamped equations
\begin{subequations}\label{eq7_1}
\begin{align}
(d_{i+1}+f) \cos \theta_i-(2d_i+f)+d_{i-1} \cos \theta_{i-1}  & = 0\label{eq7_1a}\\
(d_{i+1}+f) \sin \theta_i-(d_i+1)\omega-d_{i-1} \sin \theta_{i-1} & =0\label{eq7_1b}
\end{align}
\end{subequations}
\noindent
with $d_0=d_N+f=0$. In this work, we are focusing on the system properties in the overdamped limit.

\subsubsection{Trivial case: straight configuration}
The only rigid case where the orientation remains constant over time is when there is no resultant torque on the system i.e. when all beads are initialized on a straight line. This is the `trivial' case of unbound steady state. Each bond angle can either be $0$ or $\pi$, hence giving
$\sin \theta_i=0$, $\cos \theta_i=\sigma_i=\{+1,-1\}$ and $\omega=0$ since the motion is linear. Putting in eq. \eqref{eq7_1a}, we get
\begin{equation}\label{eq8}
\sigma_i d_{i+1}-2d_i+\sigma_{i-1}d_{i-1}=(1-\sigma_i)f
\end{equation}

\noindent



Here, for example, we discuss the `fully extended' case where $\sigma_i=1$ for all $i=1, \ldots, N-2$. Solving for $d_i$'s, we get from eq \eqref{eq8}, $d_i=-if/N$ for $i=1$ to $N$. Hence, all bonds are compressed (i.e., $d_i<0$ ) and bonds closer to the head are more compressed than those further away. This feature is even seen in other rigid steady states and also in many flexible steady states, as we discuss in the subsequent sections.

Additionally, we can calculate the steady state velocity for this configuration. For this, eq.\eqref{eq1} is summed over all $i$ and divided by $N$ on both sides. 
Noting that $\frac{1}{N} \sum_ir_i=r_c$, is the position of the COM, we get $\dot{r}_c=v_c=\left(1-\frac{1}{N}\right) f$, which gives the translational velocity of the overdamped chain. These relations have been verified numerically.

\noindent
\subsubsection{Bound state of three monomers}


\begin{figure*}[t]
    \centering
\includegraphics[width=0.7\textwidth]{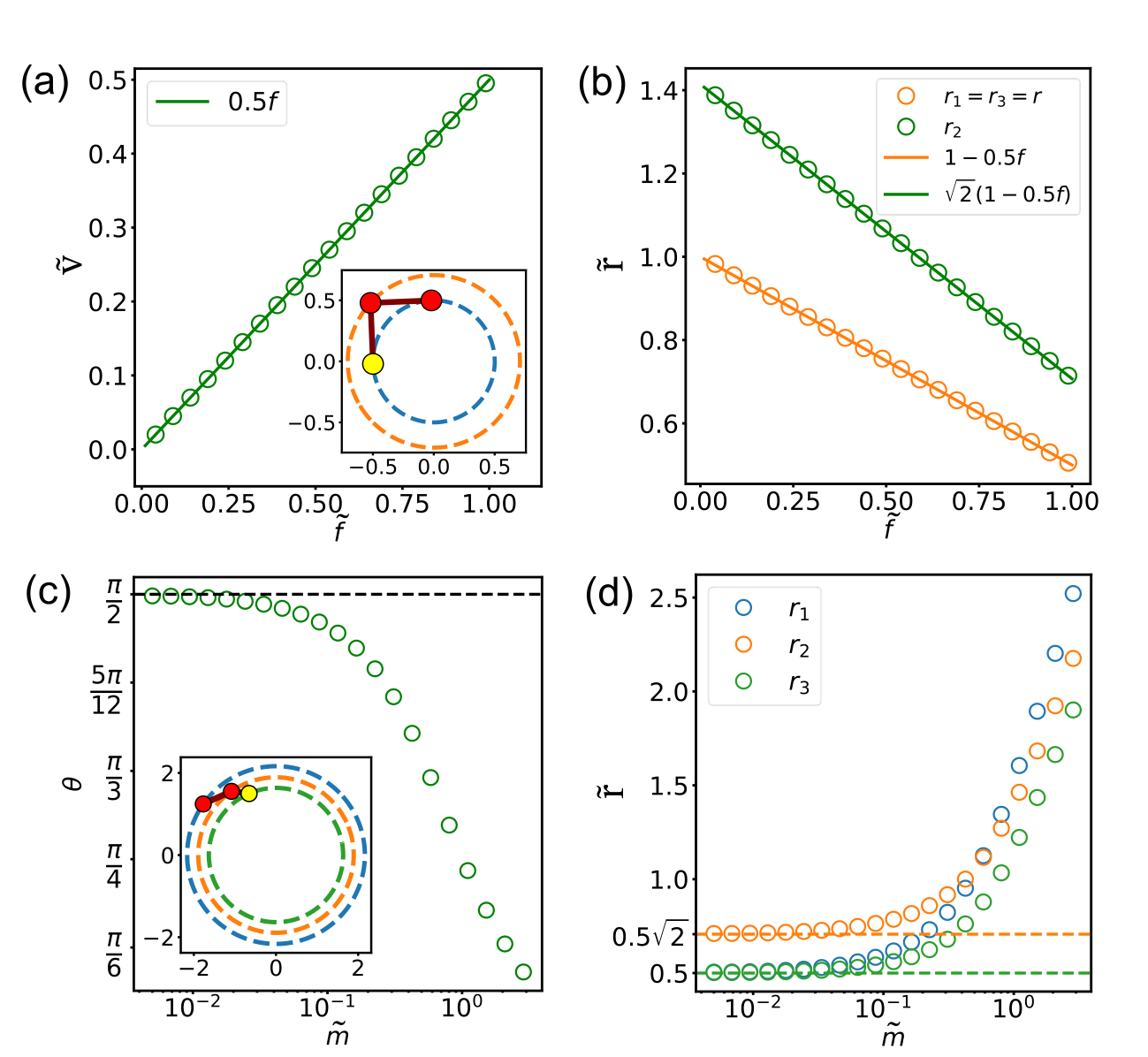} 
    \caption{(a)  Variation of dimensionless chain centre-of-mass velocity $\tilde{ V} = r\omega$ (in units of $kb/\gamma$) with non-dimensional active force $\tilde f = f/kb$, where $k$ is the spring stiffness, $\gamma$ the damping coefficient and $b$ is the equilibrium bond length. Inset: a typical steady-state configuration in the overdamped limit. (Also see MOVIE1)(b) The variation of radii $\tilde r$ (in units of $b$) of monomer trajectories with the non-dimensional active force $\tilde f$ for the 3-bead overdamped system. The first and third monomers follow the same trajectory, while the second monomer orbits with a different radius. The solid lines in (a) and (b) indicate theoretical values. (c) Variation of bond angle $\theta$ with non-dimensional mass $\tilde m=mk/\gamma^2$ for the underdamped three-bead system. As $\tilde m \rightarrow 0$, $\theta$ converges to $\pi/2$ (dotted line). Inset: a typical configuration of an underdamped three-bead system  (d) Variation monomer radii $\tilde r$ with mass $\tilde m$ for the 3-bead underdamped system for $\tilde f=1$. As $\tilde m \rightarrow 0$, the radii approaches the overdamped values (dotted lines).}
    \label{fig3}
\end{figure*}

\noindent
In this section, we specifically analyze the case of $N=3$, which is the simplest possible system that can be constructed within our framework. We can explicitly write equations \eqref{eq7_1} for $i=1,2$ as follows
\begin{subequations}\label{eq8_1}
    \begin{align}
        (d_2+f)\cos\theta-(2d_1+f)=0\label{eq8a_1}\\
        -(2d_2+f)+d_1\cos\theta=0\label{eq8b_1}\\
        (d_2+f)\sin\theta-(d_1+1)\omega=0\label{eq8c_1}\\
        -(d_2+1)\omega-d_1\sin\theta=0\label{eq8d_1}
    \end{align}
\end{subequations}
\noindent
where, we have replaced $\theta_1$ by $\theta$ and used the boundary condition $d_0=d_3+f=0$. Eliminating $\cos\theta$ from eq~(\ref{eq8a_1}), and eq~(\ref{eq8b_1}) and $\omega$ from eq~(\ref{eq8c_1}), and eq~(\ref{eq8d_1}) respectively, we get
\begin{subequations}\label{eq9}
    \begin{align}
        d_1(2d_1+f)-(d_2+f)(2d_2+f)=0\\
        d_1(d_1+1)+(d_2+f)(d_2+1)=0
    \end{align}
\end{subequations}
\noindent
Solving eq~\eqref{eq9}, we get $d_1=d_2=-\frac{f}{2}$, $0\le f\le2$ and then using \eqref{eq8_1}, we get $\theta=\frac{\pi}{2}$ and $\omega=\frac{f}{2-f}$.
Thus, in the overdamped limit, the bond lengths are equal and decrease linearly with the active force $f$. Additionally, the bond angle assumes a constant value, independent of $f$. To get the radii, we first note that the radii are simply the magnitudes of the position vectors $\bm{r}_i$, if we choose our origin to be at the center of the concentric circles traced out by the monomers, as shown in Fig \ref{fig1}(b). Since $r_i$ is fixed over time, we have $\bm{r}_i\cdot\dot{\bm{r}}_i=0$ for all $i$. Hence, taking dot product on both sides with $\bm{r}_i$, from eq (\ref{eq1}) in the overdamped limit, we get $\bm{r}_1 \cdot \hat{\bm{\Delta}}_1=\bm{r}_3 \cdot \hat{\bm{\Delta}}_2=0$. Thus, the quadrilateral $\Box O123$ in Fig \ref{fig1}(b) forms a square with $r_1=r_3=r=\Delta_1$ and $r_2=\sqrt{2} \Delta_1$ where $\Delta_1=1+d_1=1-f/2$. In Fig~\ref{fig3}(a), we plot the centre-of-mass-velocity ($r\omega$) of the system, as a function of $f$. We also plot the orbital radius of all three beads in Fig.~\ref{fig3}(b) as a function of $f$. Both these values strictly show the expected behaviour with increasing $f$. As a consistency check, we  also simulated an under-damped system (eq.~\ref{eq1}) and checked the effect of inertia by increasing $m$. As shown in Fig~\ref{fig3}(c-d), both $\theta$ and the radius $r$ saturate at values predicted by overdamped systems as $m \rightarrow 0$. Thus, in both underdamped and overdamped systems with $N=3$, the chain trajectory is bound and circular.

\subsubsection{Stability Analysis for N=3}

Going back to eq \eqref{eq5}, we again put $\bm{\Delta}_i=\Delta_i(\cos \phi_i, \sin \phi_i)^T$, and take the inner product with respect to $\hat{\bm{\Delta}}_i$ and $\hat{\dot{\bm{\Delta}}}_i$ as before, but this time without any rigidity assumption. Thus, in the overdamped limit, we get
\begin{equation}\label{eq9_1}
\begin{aligned}
    \dot{d}_{i} &= (d_{i+1} + f) \cos \theta_{i} + d_{i-1} \cos \theta_{i-1} - (2 d_{i} + f) \\
    (d_{i}+1) \dot{\phi}_{i} &= (d_{i+1} + f) \sin \theta_{i} - d_{i-1} \sin \theta_{i-1}
\end{aligned}
\end{equation}
\noindent
Hence, the rigid configuration is a fixed point of \eqref{eq9_1} and we want to check whether this fixed point is stable or unstable for the case of $N=3$. As before, we write out these equations explicitly for $i=1,2$, using appropriate boundary conditions and defining $\theta_1 = \theta$, as follows
\begin{equation}\label{eq9_2}
\begin{aligned}
    \dot{d}_{1} &= (d_2 + f) \cos \theta - 2d_1 - f\\
    \dot{d}_{2} &= d_1 \cos \theta - 2 d_{2} - f \\
    \dot{\theta} &= -\left( \frac{d_1}{d_2+1} + \frac{d_2 + f}{d_{1}+1} \right) \sin \theta
\end{aligned}
\end{equation}

Recalling for the circular steady state, $\theta = \pi/2$ and $d_1 = d_2 = -f/2$, the Jacobian matrix for the system \eqref{eq9_2} becomes,
\begin{equation*}
    J =
    \begin{bmatrix}
        -2 & 0 & -\frac{f}{2} \\
        0 & -2 & \frac{f}{2} \\
        \frac{4(f - 1)}{(f - 2)^2} & -\frac{4}{(f - 2)^2} & 0
    \end{bmatrix}
\end{equation*}
\noindent
and the corresponding eigenvalues are
\begin{equation*}
    \lambda = -2,\quad -1 \pm i \frac{\sqrt{f^2 + 4f - 4}}{f - 2}
\end{equation*}

Note that the first eigenvalue is always negative. The other two eigenvalues are complex with negative real parts for $f > 2\sqrt{2} - 2$, and real negative for $0 < f \leq 2\sqrt{2} - 2$. We have verified that this configuration is stable for any $f > 0$ using simulations.

In the same spirit, it can be shown that for the two linear cases for $N = 3$, namely
$\{ d_{1} =  - f / 3,\ d_{2} =  - 2f / 3,\ \theta = 0 \}$ with $0<f<1.5$ and $\{ d_{1} =  - f,\ d_{2} = 0,\ \theta = \pi \}$ with $0<f<1$, we get the eigenvalues as
\[
\left\{ \frac{f^2}{9 - 9f + 2f^2},\ -1,\ -3 \right\} \quad \text{and} \quad \left\{ -1,\ -3,\ \frac{f^2}{1 - f} \right\}
\]
respectively. These solutions are stable only in the range $1.5 \leq f \leq 3$ and $f \geq 1$, respectively. However, since bond lengths, $\Delta_i = 1+d_i$, decreases with $f$, this stable range does not correspond to any physically realizable state, because the bond length(s) become negative for $f>1.5$ in the first case and for $f>1$ in the second case. Therefore, only the circular state is stable in the entire domain of $0<f<2$.

\begin{figure*}[t]
    \centering
\includegraphics[width=\textwidth]{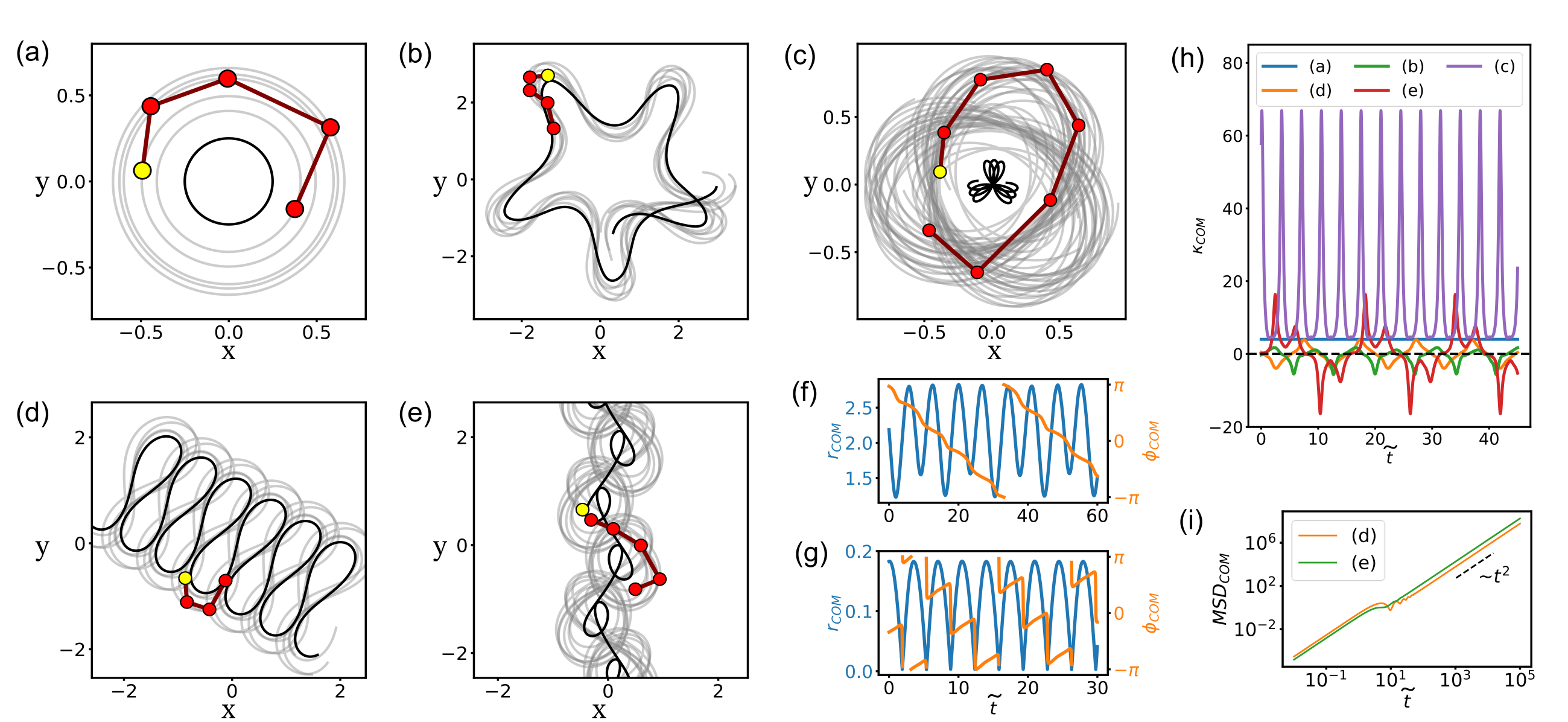}
    \caption{(a-e) Some examples of the  trajectories obtained for $N\ge4$ for the overdamped system with 2D initialization (Also see MOVIE2 - MOVIE6). The individual monomer trajectories are shown in gray and the centre-of-mass(COM) trajectory is shown in black color, obtained over a brief time interval. The head monomer (passive) is marked by yellow color. Of these, trajectories (a-c) are bounded, corresponding to $N=5$ for two different initializations and $N=8$ respectively and trajectories (d-e) are unbounded, corresponding to $N=4$ and $6$ respectively. (f-g) Variation of radius and angle subtended by the COM corresponding to the bounded trajectories (b-c) respectively. (h) Variation of the signed curvature $\kappa$ of the COM corresponding to the trajectories (a-e). (i) Mean Square Displacement(MSD) of the COM for the wavelike trajectories (d-e), which goes ballistic in long term. In figures (f-i), $\tilde t$ represents the dimensionless time in units of $\gamma/k$.}
    \label{fig4}
\end{figure*}

Unfortunately, the analysis we did in this section does not generalize to all $N$ in the Overdamped limit. In general, for higher values of $N$, the steady state configurations are sensitive to initial conditions and the chain segments trace a rather complex trajectory in many cases, as we will discuss in the subsequent section. These dynamical states could not be analyzed using analytical methods. Therefore, we employ numerical methods to study the general case. However, for very large values of $N$, we recover a general pattern again, which we have discussed later.
\subsection{Numerical analysis of $N > 3$}
One of the striking features of this system is the variety of patterns made by the particle trajectories in systems with $N>3$. These patterns depend on both $N$ and the initial configuration of the chain.
Some examples of these trajectories are shown in Fig~\ref{fig4}. For $N=5$ (see MOVIE3 and MOVIE4)   (Fig~\ref{fig4}(a)-(b)) and $N=7$, we observe bound trajectories, whereas for $N=4,6$ we the system displays unbound trajectories (Fig~\ref{fig4}(d)-(e)) and (see MOVIE2 and MOVIE5). For $N \geq 8$, the chain always follows bound trajectories (Fig~\ref{fig4}(c)) (also see MOVIE6). 
The configuration of the chain following can be rigid (Fig~\ref{fig4}(a)), leading to a periodic and circular trajectory. The examples are $N=5,7$. In some other cases, the configuration is also flexible (Fig~\ref{fig4}(b)-(c)). These trajectories resemble complex spirograph patterns~\cite {WikipediaSpirograph2025}. Also, the centre-of-mass trajectories displayed by chains appear to be quasi-periodic as they never repeat and densely cover a region in the x-y plane. 
\subsubsection{Quantification of steady states}
To analyse the dynamical patterns displayed by the chains more quantitatively, we compute the signed curvature ($\kappa$) of the trajectories traced out by the centre-of-mass (COM) of the polymer in time. In Fig~\ref{fig4}(h), we plot these values of $\kappa$ against time for different trajectories. 
The time evolution of 
For rigid, bound states (Fig~\ref{fig4}(a)), the periodic, circular trajectory ensures a non-zero $\kappa$, invariant in time. For flexible bound states with quasiperiodic trajectories (Fig~\ref{fig4}(b)-(c)) $\kappa$ is not invariant in time. However, these variations are asymmetric about zero, hence provide a non-zero time-averaged value $\bar\kappa$. For unbound states (Fig~\ref{fig4}(d)-(e)), the COM trajectories show periodic oscillations. However, at a much larger timescale, the COM moves along a straight line. Therefore, short-time oscillations in $\kappa$ are symmetric about zero, hence, $\bar\kappa =0$. 

Although the centre-of-mass of the chain follows quasiperiodic trajectories for $N=5$ and $N=8$, they are qualitatively different, as evident in Fig~\ref{fig4}(b) and \ref{fig4}(c). To analyse these trajectories closely, we switch to polar coordinates with the origin at the centre of symmetry. In Fig~\ref{fig4}(f) and Fig.~\ref{fig4}(g), we plot both raidal ($r_{COM}$) and the angular ($\phi_{COM}$) coordinates of the trajectory, in time. It is evident that $r_{COM}$ shows a period 2 behaviour for $N=5$ and a period-1 behaviour for $N=8$. Further, the amplitudes of these oscillations are smaller for $N=8$, indicating an enhanced confinement in that case. Both these trajectories are also characterised by periodic undulations in $\phi_{COM}$, about a constant angular velocity (Fig~\ref{fig4}(g)-(f)), where one continuous curve indicates one complete rotation. The $\phi_{COM}$ amplitudes are more pronounced for $N=8$. The number of such oscillations within a complete rotation is an indicator of the number of lobes in a bound trajectory. The periodicity of these undulations is incommensurate with the centre-of-mass orbital period in both cases. This incommensurability leads to quasiperiodicity in these trajectories.

On the other hand, unbound trajectories show a periodic behaviour as they steadily translate in space. Such wave-like trajectories are only possible for flexible configurations, where the bond angles periodically vary in time. For $N=4$ such trajectories follow a regular meandering pattern (Fig~\ref{fig4}(d)), whereas for $N=6$ they display periodically looping trajectories (Fig~\ref{fig4}(e)). In Fig~\ref{fig4}(i) we compute the mean-square deviation (MSD) in both these cases, calculated using a moving-window average for long-time trajectories. It is evident that for time-scales larger than the periodicity, the COM display a ballistic motion, as the MSD scales as $t^2$

\subsubsection{Dependence on Initial condition}

\begin{figure}[h]
    \centering
\includegraphics[width=.5\textwidth]{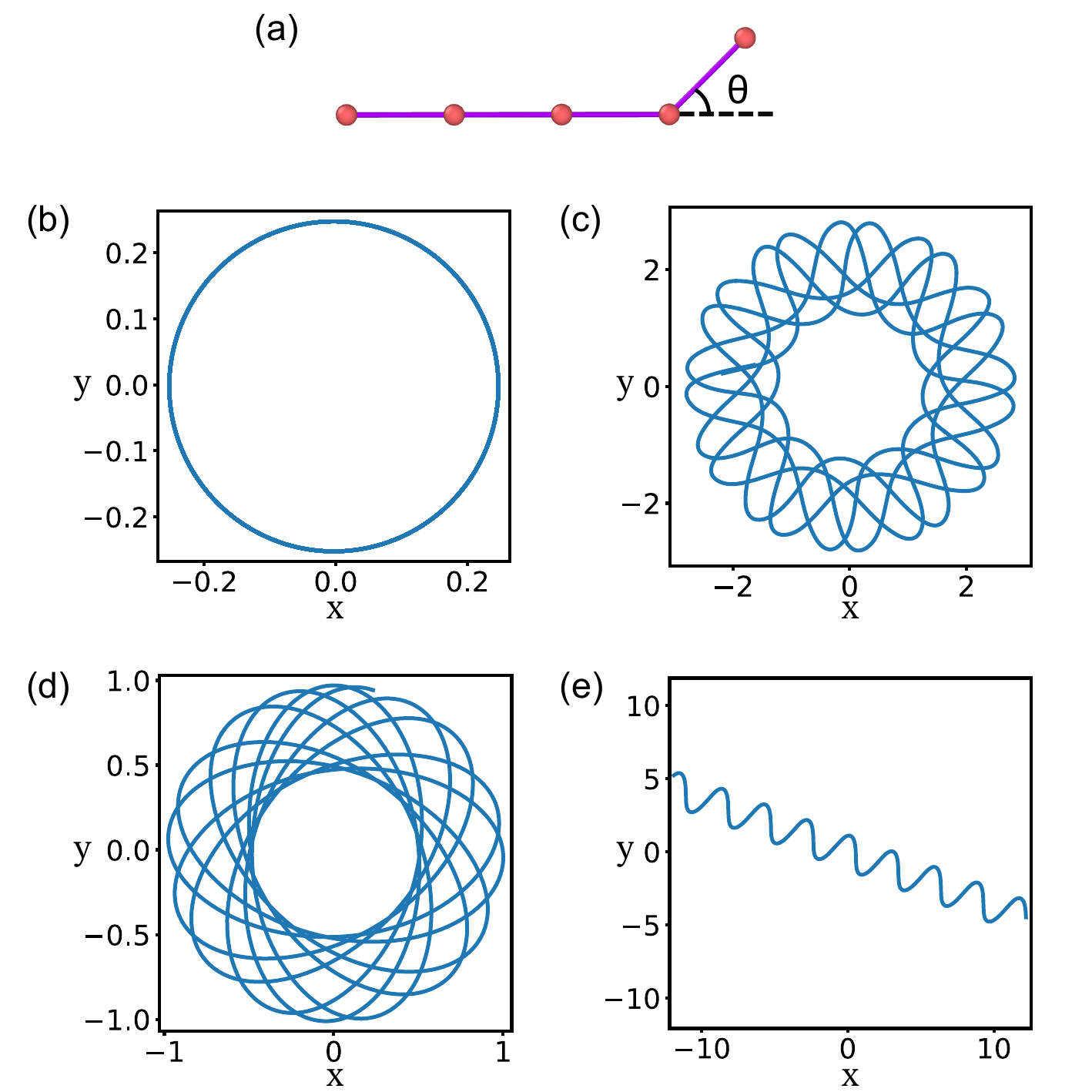}
    \caption{Dependence of the steady state trajectory for some values of $N$ and $f$ started from different initializations generated by varying angle $\theta$ as in (a). Parameter values are as follows. (b-c) $N=5$, $f=1$, (b) $\theta=3\pi/5$ and (c) $\theta=4\pi/5$. (d-e) $N=7$, $f=0.5$, (d) $\theta=4\pi/5$, (e) $\theta=\pi$}
    \label{fig5}
\end{figure}

Another interesting feature of this system is that there is a dependence on the initial conditions of the steady state. This suggests that possibly there exists more than one steady state, often of different varieties, each having its own basin of attraction. To illustrate this property, we present two such instances of initialization dependence. We initialize the system from a straight configuration $(\theta_i =0)$, except for the last bond angle $\theta_{N-2}$. More specifically, we choose $\Delta_i=1,i=1,\dots,N-1$; $\theta_i=0,i=1,\dots,N-3$ and $\theta_{N-2}=\theta$, as shown in Fig. \ref{fig5}(a). With this initialization, for $N=5$ we observed that for $\theta=3\pi/5$ the system goes to a rigid bound state (Fig. \ref{fig5}(b)) whereas for $\theta=4\pi/5$ the system goes to a flexible bound state (Fig. \ref{fig5}(c)). Changes in initial condition can also change a bound steady state to unbound, for example, Fig. \ref{fig5}(d-e) shows the the initializations for $N=7,f=1/2,\theta=4\pi/5$ and $\theta=\pi$ respectively. The former gives a bound steady state, whereas the latter gives an unbound steady state. However, if we initialize the chain from a random coiled configuration, the chain always goes to a rigid bound state as in Fig.~\ref{fig5}(b), indicating that it is the highest probable steady state for this system.\\



\subsection{Large $N$ limit}

\begin{figure}[h]
    \centering
    \includegraphics[width=.5\textwidth]{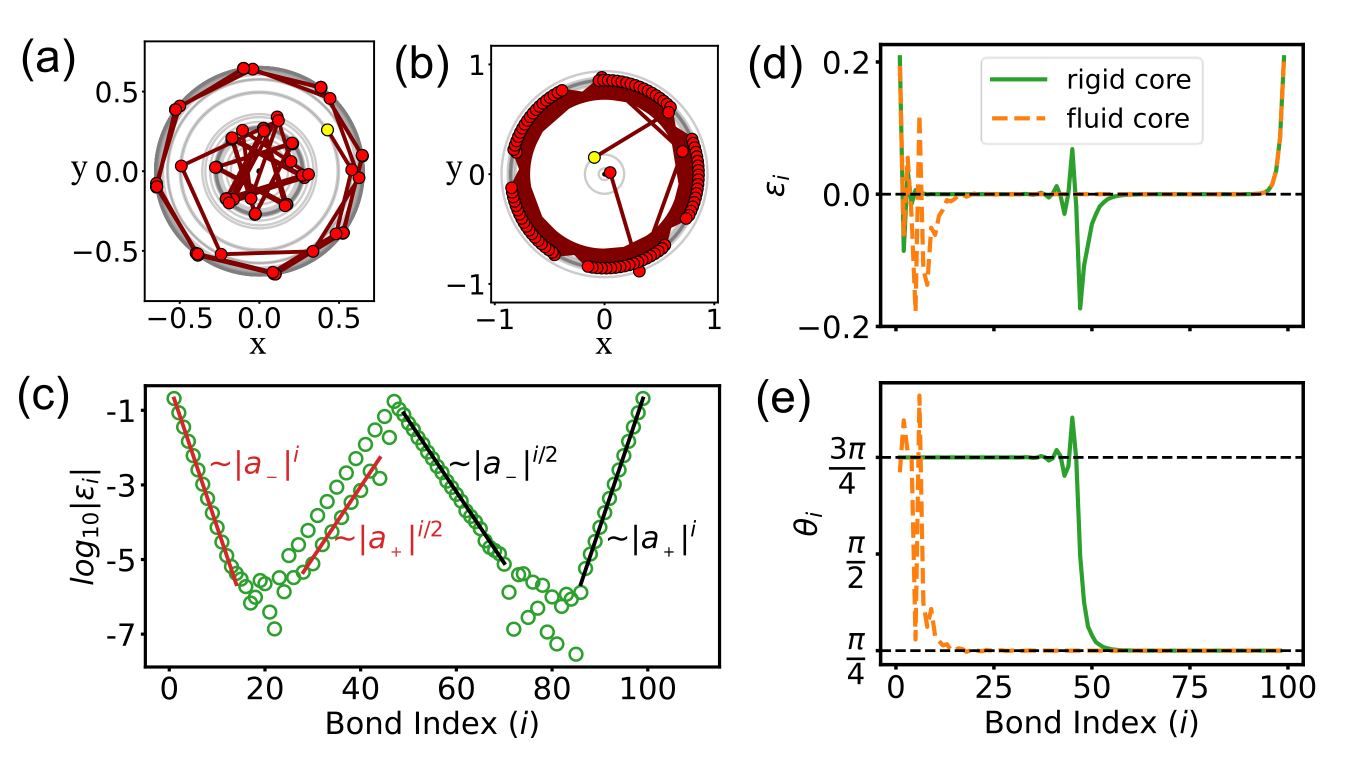}
    \caption{(a-b) The circular trajectories observed, along with the configuration of a chain with the number of segments, $N=100$, for two types of tangent definitions. (a) The tangent vector is same as the bond vector (b) the tangent is the average of two consecutive bond vectors. (c) Variation of the log absolute deviation ($\log_{10}|\varepsilon_i|$) with bond index $i$ for the chain shown in (a).  Circles are the simulated data points, and the solid lines are drawn to compare the slopes in log scale to the theoretical predictions. The red lines correspond to the tail solution and the black lines to the head solution. (d-e) Variation of deviations $\varepsilon_i$ and bond angles $\theta_i$ respectively with bond index. The solid green line corresponds to trajectory (a), the dashed orange line to another type of trajectory with a flexible core for comparison. The black dashed lines corresponds to theoretical predictions for the saturating values.}
    \label{long_chain}
\end{figure}

In the previous sections, we saw that simulating the overdamped system for $N>3$ leads to all kinds of complex steady-state trajectories, which simply tells us that there exist multiple different types of steady-states with possibly different basins of attraction. However, when we looked at the large $N (\gtrsim 30)$ behaviour for $f=1$, the numerical studies show that the polymer folds into a circular steady state with a significant fraction of the chain having constant radius, bond length and bond angle (Fig~\ref{long_chain}(a)) (also see MOVIE7). These configurations lead to self-intersecting loops, as the loop size is smaller than the chain length. To avoid the self-interaction, typically excluded volume interactions between the chain segments are added. However, such interactions introduce randomness in the system, which eventually leads to the formation of a coiled state, as observed in the previous studies~\cite{bianco2018globulelike, malgaretti2025coil}. Since our goal is to understand the dynamical states of the system in the absence of any noise, we do not introduce any excluded volume interactions and allow the chains to intersect in the looped state. Although these states are unrealistic for real chains, our analysis reveals many interesting aspects from the perspective of an active chain as a dynamical system and helps to understand the origin of periodic behaviour observed in many realistic models of tangentially driven chains. 

A more careful inspection reveals that bond lengths of the chain remain constant for most part of the chain, and the bond angle shifts from one value to another, along the chain contour, as shown in Fig~\ref{long_chain}(d)-(e). These regions determine the overall dimensions of the resulting trajectory and the loop, hence demand a thorough investigation.
As a first step, we consider the region where the bond lengths and bond angles are independent of the bond index, denoted by $\Delta$ and $\theta$ respectively. Let us also define $d=\Delta-1$ as the bond extension which is independent of the bond index. Putting in eq \eqref{eq7_1} and solving, we get the relations,
\begin{subequations}
\begin{align}
d &= -\frac{f}{2} \label{eq12_n_1}\\
\omega (d+1) &= f\sin\theta \label{eq12_n_2}
\end{align}
\label{eq12_n}
\end{subequations}
 To proceed further, we consider the region where the bond length deviates from $\Delta$ while keeping $\theta$ constant. We define this deviation $\varepsilon_i=\Delta-\Delta_i=d-d_i$. Substituting this in
 eq.\eqref{eq7_1} and using eq.~\ref{eq12_n}, we get the relations,
\begin{equation}\label{eq10}
\begin{aligned}
\varepsilon_{i+1}+\varepsilon_{i-1}&=2\varepsilon_i\sec\theta\\
\varepsilon_{i+1}-\varepsilon_{i-1}&=\omega\varepsilon_i\text{ cosec }\theta.
\end{aligned}
\end{equation}
\noindent
with $\varepsilon_0=-\varepsilon_N=d$, to account for the boundary conditions of \eqref{eq7_1} in terms of $\varepsilon$. The eq \eqref{eq10} indicates a solution of the form $\varepsilon_i=a^i\varepsilon_0$, where $a=\sec\theta-\frac{d}{d+1}$, and $a^{-1}=\sec\theta+\frac{d}{d+1}$ and taking their product we get, $\sec^2\theta=1+(\frac{d}{d+1})^2$. Subsequently, we get two values of $\theta$ as $\theta=\alpha$ and $\pi-\alpha$, where $\alpha=\tan^{-1}(-\frac{d}{d+1})=\tan^{-1}(\frac{f}{2-f})$. Note that the possible range of $\theta$ is fixed to $[0,\pi]$, since $[-\pi,0]$ simply represents the same configuration rotating in the opposite direction. In the physically valid range of $f\in(0,\infty)$, $\alpha\in(0,\frac{\pi}{2}]$. For $f=1$, the two steady-state values of $\theta$ are given by $\pi/4$ and $3 \pi/4$, which have been verified by simulations, as plotted in Fig. ~\ref{long_chain}(e).

Note that the particular form of solution for $\epsilon_i$ cannot simultaneously satisfy both boundary conditions. Hence, we need to satisfy the two boundaries separately as $\varepsilon_i=a^i\varepsilon_0$ near the tail and $\varepsilon_{N-i}=a^{-i}\varepsilon_N$ near the head, and both these $a$'s cannot be the same. But, one can get two soultions to $a$ from two values of $\theta$, which are $a_\pm=\tan\alpha\pm\sec\alpha$ for $\theta =\alpha, \text{and }\pi-\alpha$. For the range $\alpha\in(0,\frac{\pi}{2}]$, we obtain the possible range of $a_\pm$,  $a_+\in(1,\infty)$, and $a_-\in(-1,0]$ (or $a_+^{-1}\in[0,1)$, $a_-^{-1}\in(-\infty,-1)$). Considering the condition that $\epsilon$ is finite and $|a| \leq 1$ for all $i$,  we assign $a_{-}$ to the tail end and $a_{+}$ to the opposite end.
Noting that $\varepsilon$ diverges for $|a|>1$, we obtain $\varepsilon_i=a_-^i\varepsilon_0$ and $\varepsilon_{N-i}=a_+^{-i}\varepsilon_N$. 

However, the above assumption of constant $\theta$ is only valid near the ends of the chain. Since the $\theta$ values are different at each end, there will be a mismatch towards the middle segments of the chain where these values crossover.
To understand the correct structure at the interior of the chain, we redo the same analysis, but this time assuming deviations in $\theta$, such that $\eta_i=\theta-\theta_i$. We are assuming small $\eta_i$s, which is valid in the immediate vicinity of the constant $\theta$.  
Considering again the deviations at the bond lengths $\varepsilon_i=d-d_i$ and substituting these in eq \eqref{eq7_1}  we get,
\begin{equation}\label{eq11}
\begin{aligned}
    (\varepsilon_{i+1}+\varepsilon_{i-1})\cot\theta-2\varepsilon_i\text{ cosec }\theta&=-d\eta_i+d\eta_{i-1}\\
    (\varepsilon_{i+1}-\varepsilon_{i-1})\tan\theta-\omega\varepsilon_i\sec\theta&=d\eta_i+d\eta_{i-1}
\end{aligned}
\end{equation}
with $\varepsilon_0=-\varepsilon_N=d$ as before and $\eta_0=\eta_{N-1}=0$. Note that we have retained only the terms only upto the first order in both $\varepsilon_i$ and $\eta_i$. 
Rearranging the terms, we can write from eq \eqref{eq11},
\begin{equation}\label{eq12}
\begin{aligned}
    2d\eta_i=\pm A_-\varepsilon_{i+1}\mp A_+\varepsilon_{i-1}-2B_\pm\varepsilon_i\\
    2d\eta_{i-1}=\pm A_+\varepsilon_{i+1}\mp A_-\varepsilon_{i-1}+2B_\mp\varepsilon_i,
\end{aligned}
\end{equation}
where $A_\pm=\tan\alpha\pm\cot\alpha$ and $B_\pm=\pm\tan^2\alpha-\text{ cosec }\alpha$. Note that $+$ and $-$ symbols correspond to head and tail ends, respectively. Equating and eliminating $\eta_i$ from \eqref{eq12}, we get
\begin{equation}\label{eq13}
    \pm A_+\varepsilon_{i+2}+(2B_\mp\mp A_-)\varepsilon_{i+1}+(2B_\pm\mp A_-)\varepsilon_i\pm A_+\varepsilon_{i-1}=0
\end{equation}
which has to be satisfied for all $i$. To solve this, we try an ansatz of the form $\varepsilon_i \sim a^i$ in eq \eqref{eq13} which gives us a cubic equation in $a$. However, we already know that $a_\pm=\tan\alpha\pm\sec\alpha$ as obtained before must be a solution for $a$, because eq \eqref{eq13} is satisfied even when $\eta_i=0\;\forall i$ in \eqref{eq12}. Factoring out this trivial root from the resulting cubic equation in \eqref{eq13}, we get that the other two roots must be roots of the following quadratic equation
\begin{equation}\label{eq14}
    a^2+(\cos{2\alpha})(1+a_\mp)a+a_\mp=0
\end{equation}

Let the roots be $b_\pm$ and $c_\pm$, which can be easily calculated. Then, the general solution for $\varepsilon_i$ is of the form $\varepsilon_i=p_\pm a_\pm^i+q_\pm b_\pm^i+r_\pm c_\pm^i$ for the head and tail sides respectively. There are six undetermined constants, two of which are determined from the boundary conditions $\varepsilon_0=-\varepsilon_N=d$ and the other four has to come from the interior part where the left and right solutions meet. \\
For a specific case with $f=1$, we get $\alpha=\frac{\pi}{4}$, $a_\pm=1\pm\sqrt2$. The other two roots are obtained from eq \eqref{eq14} as $b_\pm=\sqrt{-a_\mp}$ and $c_\pm=-\sqrt{-a_\mp}$. Note that these can be complex, but the coefficients will be such that the final expression for $\varepsilon_i$ is real. So, the solution for $\varepsilon_i$ towards the head is given by $\varepsilon_i=p_+ a_+^i+Q_i^+ (-a_-)^{i/2}$ and towards the tail is given by, $\varepsilon_i=p_- a_-^i+Q_i^- (-a_+)^{i/2}$. In addition, the boundary condition insists that $p_++Q_0^+=\varepsilon_0=-0.5$ and $p_- a_-^N+Q_N^-(-a_+)^{N/2}=\varepsilon_N=0.5$, where $Q_i^\pm=q_\pm+(-1)^ir_\pm$. To compare with the simulations, we plot $\log|\varepsilon_i|$ in Fig.~\ref{long_chain}(d) for the case of $N=100$, which confirms that the solution behaves like $|a_\pm|^i$ near the ends and like $|a_\pm|^{i/2}$ at the interior (Fig~\ref{long_chain}(d)). The simulations also reveal that the exact point where the left and right solutions meet is not fixed, but varies in different realisations starting from different initial conditions.\\
One can also calculate the angles $\theta_i$ from $\eta_i$ in terms of $\varepsilon_i$ from eq \eqref{eq12}, as well as the angular velocity of the chain, given by $\omega=\frac{f\sin\theta}{d+1}$(from \eqref{eq12_n_2}). From bond angle $\theta$ and the bond length $\Delta$, we can also calculate the effective radius of the loop given by $r=\frac{d+1}{2}\text{cosec}(\frac{\theta}{2})$ for any general $f$. Since the segments near the two ends that define the geometry of the loop, we can use the constant $\theta$ values near both ends for calculating the loop radius. Using this, one can obtain the radii at both ends as $r_\pm=\frac{d+1}{\sqrt{2(1\mp\cos\alpha)}}$ where ($+$) denotes the head and ($-$) represents the tail. Since $\cos\alpha$ is always positive for $\alpha\in(0,\frac{\pi}{2}]$, we have $r_+>r_-$. For example, putting  $f=1$, $d=-1/2,\text{ and }\alpha=\pi/4$, we obtain $r_\pm=\frac{1}{2\sqrt{2\mp\sqrt{2}}}$. This difference in radii manifests as a smaller core of radius $r_-$ within a larger loop of radius $r_+$ as seen in Fig \ref{long_chain}(a). We have termed this configuration as `rigid core' as the configuration shows a constant radius at the inner side of the loop. Of course, the number of monomers in the inner core and the outer loop may vary. However, in some of the realisations we saw the tail monomers move in irregular trajectories, whereas the opposite end forms a well-defined circular loop. Such realisations are termed `fluid core' to contrast with rigid core solutions. The differences in bond lengths and angles between these two types of solutions is highlighted in Fig \ref{long_chain}(d) and (e). 

The above analysis is valid for a sufficiently large $N$, for which a region of constant bond length and bond angle can exist. We now estimate the minimum chain length required for the existence of such a regions. 
As noted earlier, the deviations $\varepsilon$ decay as $\varepsilon_i\sim a^i$ at any section of the chain. This gives us a typical length scale of decay, $\lambda$ such that $\varepsilon_i\sim e^{i/\lambda}$. Hence, $\lambda=\frac{1}{|\log a|}$. For $f=1$, we have already shown that $|\varepsilon_i|\sim|a_\pm|^i$ near the ends and $|\varepsilon_i|\sim|a_\pm|^{i/2}$ at the interior. Adding all four contributions, we get the total length scale $\lambda=\frac{1}{|\log{|a_-|}|}+\frac{2}{|\log{|a_+|}|}+\frac{2}{|\log{|a_-|}|}+\frac{1}{|\log{|a_+|}|}=\frac{6}{\log a_+}\approx6.81$ for $f=1$. Consistent with this estimation, in our simulations, we never see spatially unbound, wavelike solutions (Fig. \ref{fig4}(d-e)) for chains longer than $\lambda$. Hence, $\lambda$ determines the minimum chain length required to obtain a large $N$ dynamics.

\subsection{Alternate tangent definition}

We have also explored, in our simulations, an alternate definition of the `local tangent', that has been used previously in many studies~\cite{bianco2018globulelike}. According to this, the unit tangent vector is defined as $\hat{\bm{p}_i}= (\bm{r}_{i+1} - \bm{r}_{i-1})/|\bm{r}_{i+1} - \bm{r}_{i-1}|$. 
With this tangent definitions, the polymer is passive at both ends. We have simulated the overdamped chain dynamics  with this tangent definition, for a range of initial conditions, for $f=1$.
Numerical simulations show that even for small $N$, the chain often forms highly folded configurations with angles close to $\pi$.
However, in the large $N$ limit, we observe rigid, circular configurations (Fig \ref{long_chain}(b)) similar to the former tangent definition (Fig \ref{long_chain}(a)).
This property indicates that circular states in the large $N$ limit are a generic feature of these systems, irrespective of the specific tangent definition used.
\subsection{3D Initializations}

\begin{figure}[h]
    \centering
    \includegraphics[width=.5\textwidth]{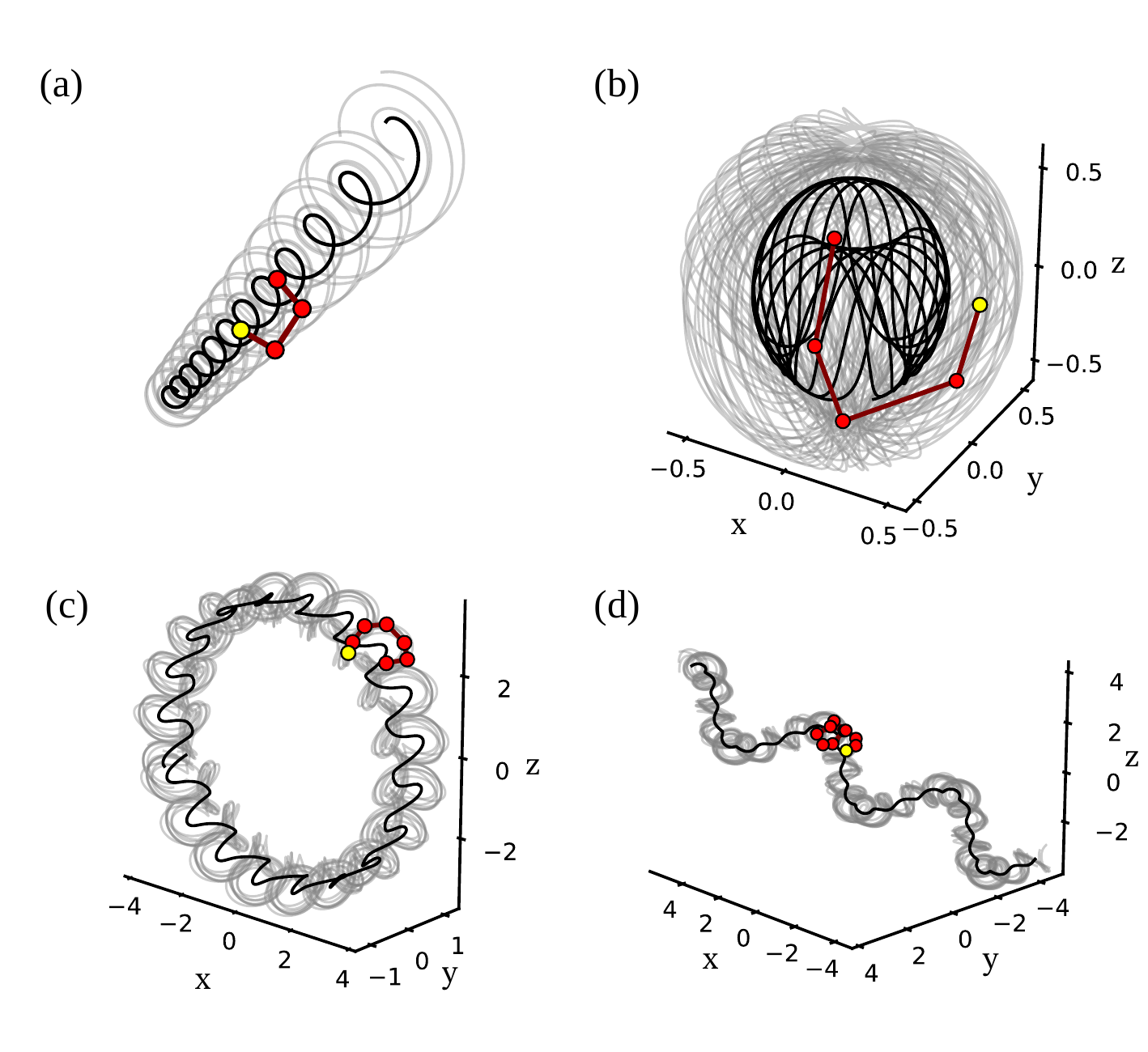}
    \caption{Trajectories obtained for some 3D initial conditions: (a) $N=4$, (b) $N=5$, (c) $N=7$ and (d) $N=10$. (a,d) are unbound, and (b,c) are bound. Also see MOVIE8- MOVIE11.}
    \label{fig8}
\end{figure}

Although the equations of motion for the chains are in three dimensions, their trajectories are confined to a plane so far, since the initial chain configurations are planar. To explore the out-of-plane trajectories, we initiate the chains with random coiled configurations in three dimensions. Interestingly,  
 these systems add even more richness to the patterns formed by steady state trajectories (see MOVIE8- MOVIE11). Some examples are given in Fig~\ref{fig8}. We observe many more instances of unbound trajectories than bound ones. They include regular helix (Fig~\ref{fig8}(a), $N=4$) and super-coiled helix (Fig~\ref{fig8}(d), $N=10$). We also observe spatially bound steady states, which include globular (Fig.~\ref{fig8}(b), N=5) and wavy ring (Fig.~\ref {fig8}(c), N=7).  
 Interestingly, for large $N$, the chain forms ring-like structures similar to (Fig \ref{long_chain}(a)) with of similar radius. However these structures do not stay stable for the entire simulation time. Instead, they tend to dissolve away from circular configurations and re-form cyclically during the dynamics. The cyclic structures are not entirely planar, since they show finite width perpendicular to the radial direction. We have also studied the system with predominantly straight configuration except for the last two bond angle where we impose non-planar deviation and we obtain similar results as random initializations.

\section{Summary and Outlook}

This work explores the dynamics of a collection of particles linked by an elastic potential with local interactions defined by follower activity. The number of segments varies from the minimum value, $N=3$ to $N=200$. We have shown, both analytically and numerically, that the centre-of-mass of the chain follows a stable circular trajectory in the overdamped limit, with radius and angular velocity determined by the magnitude of the follower force. For intermediate chain lengths, the system displays rich dynamical states, as the centre-of-mass follows complex trajectories, both bound and unbound in space. For a sufficiently large segment length, the system shows a length-independent behaviour, as the chain conforms into a circular shape and its centre-of-mass follows a circular trajectory. In this limit, we analytically calculate the variation in bond length and bond angle along the chain, the parameters that determine the chain conformation. These results have been verified numerically, in the overdamped limit. In addition, we numerically explore the conformational and dynamical states displayed by these chains in 3-D. The results once again reveal a rich class of dynamical states in 3D. 

Since the aim of our study is to understand the interplay between active and elastic forces, we have simplified the system by ignoring the excluded volume interactions and thermal fluctuations. It is known that when the noise level is significant, the flexible chain conformation becomes a random coil. Although the excluded volume does not affect the chain conformation and chain dynamics at relatively small chain lengths, for longer chains, it becomes a source of randomness, which again leads to coiled states. However, these simplifications allow us to understand the origin of the periodic behaviour observed in a range of realistic models, such as clamped active filaments~\cite{chelakkot2014flagellar, de2017spontaneous, ling2018instability, Fatehiboroujeni2018, fily2020buckling, sangani2020elastohydrodynamical} and motor-driven filament assays~\cite{shee2021semiflexible, yadav2024wave, ng2025active, collesano2022active}. Our study reveals the minimal requirement for producing periodic behaviour in connected, active systems. 

In addition, these findings will be potentially relevant in the context of developing and designing robotic system ~\cite{7hcf-p1yk, paramanick2024uncovering, son2025emergent}, as we show the existence of a wide variety of locomotive modes in 2D and 3D, by changing various system parameters. Moreover, our study reveals the emergence of spatiotemporal coordination arising from nearest-neighbor interactions. These dynamical states could further be enriched by introducing internal variables, such as phases, that can be synchronized among connected entities and couple to their activity, as explored in recently developed swarmalator models~\cite{o2017oscillators, sar2025swarmalators, kreienkamp2025synchronization}.

\begin{appendix}
    {\bf Appendix: Supplementary movies}\\
    MOVIE 1: Circular trajectory observed for $N=3$ \\
    MOVIE 2: Unbound periodic trajectory observed for $N=4$ \\
    MOVIE 3: Circular trajectory for $N=5$\\
    MOVIE 4: Non-circular bound trajectory (spirographic pattern) for $N=5$ with alternate initial condition \\
    MOVIE 5: Complex, unbound trajectory of the chain for $N=6$\\
    MOVIE 6 : Non-circular, bound trajectory (spirographic pattern) observed for $N= 8$\\
    MOVIE 7: Large $N$ behaviour -- circular trajectory observed for $N=100$.\\
    MOVIE 8: 3D intial conditions for $N=4$. The chain displays a helical trajectory.\\
    MOVIE 9: Bound motion for $N=5$ with 3D initial conditions.\\
    MOVIE 10: Supercoiled bound trajectory observed for $N=7$.\\
    MOVIE 11: Supercoiled unbound trajectory observed for $N=10$.
\end{appendix}

\begin{acknowledgments}
\noindent
We thank Amitabha Nandi for helpful discussions. SS thanks Pampa Dey for helping towards the preparation of the manuscript.
\end{acknowledgments}

\bibliographystyle{apsrev4-2}
\bibliography{Citation-Cluster}

\end{document}